# Impact of International Cooperation for Sustaining Space-Science Programs

## Karan Jani[1]

*Sam Nunn Security Program*
Sam Nunn School of International Affairs
Georgia Institute of Technology
October 26, 2016

## ABSTRACT

Space-science programs provide a wide range of application to a nation's key sectors of development: science-technology infrastructure, education, economy and national security. However, the cost of sustaining a space-science program has discouraged developing nations from participating in space activities, while developed nations have steadily cut down their space-science budget in past decade. In this study I investigate the role of international cooperation in building ambitious space-science programs, particularly in the context of developing nations. I devise a framework to quantify the impact of international collaborations in achieving the space-science goals as well as in enhancing the key sectors of development of a nation. I apply this framework on two case studies, (i) *Indian Space Research Organization* - a case of space-science program from a developing nation that has historically engaged in international collaborations, and (ii) *International Space Station* - a case for a long term collaboration between matured space-science programs. My study concludes that international cooperation in space can significantly enhance scope of science goals, but has relatively little return of investment towards science education and national security. I also highlight limitations of such space diplomacy in the case of China and SAARC nations, and list criteria for future investigation and case studies.

[1] kpj@gatech.edu

# TABLE OF CONTENTS

*"Yours will be a future where human beings have pushed farther into the universe, not just to visit but also to stay. To me, public diplomacy and (international) cooperation in space go together like peanut butter and jelly... When we go up to cislunar space, it's going to give our international partners an opportunity to be with us, because no venture into deep space is going to be done by one nation...* **The Future Of Space Policy Is Built On International Cooperation***"* [2]

-**Charles Bolden,** current Administrator of NASA, retired United States Marine Corps Major General, and former NASA astronaut

# 1. Introduction

## 1.1 Historical Developments in Space Research

The love story between humans and space has ancient origins. Astronomy is placed as the oldest form of natural science, with conclusive evidence dating as back as 1600 BC. However, the idea of "exploring space" is fairly recent adventure. After the Cavendish Experiment in 1797 determined the mass of the earth, the escape-velocity to leave earth's gravity was rather well calculated 40,000 km/h. But to develop any engineering device that can have a projectile motion of this speed remained an unexplored research territory, until early 1900s, when Robert H. Goddard published seminal works[3] on liquid-fuel rockets. The scientific motivation for Goddard was to take pictures of the earth from the atmosphere. But skepticism of the academic society at the time and scarcity of funding failed to gain him much success on that route. It was not until two years after Goddard's death, in 1946, when the first picture of earth was taken by the guided ballistic missiles *V-2* that were developed by the United States during World-War-II. Following this all of mankind's attempt to explore and understand space has been a by product of military research.

From a defense perspective, the big application of *V-2* missiles was its ability to survey large area of the earth from space. This led the USSR to aggressively pursue its Soviet Space Program. By late 1950s, the Soviets developed the first intercontinental missile *R-7 Semyorka*, which had enough energy to cross escape velocity of earth and placed the first manmade object in space to orbit the earth *- Sputnik-1*. This build the pressure on the US to enter the "space-race" and in 1958 President Eisenhower instituted the National Aeronautics and Space Administration (NASA). Unlike the Soviet Space Program, the US

---

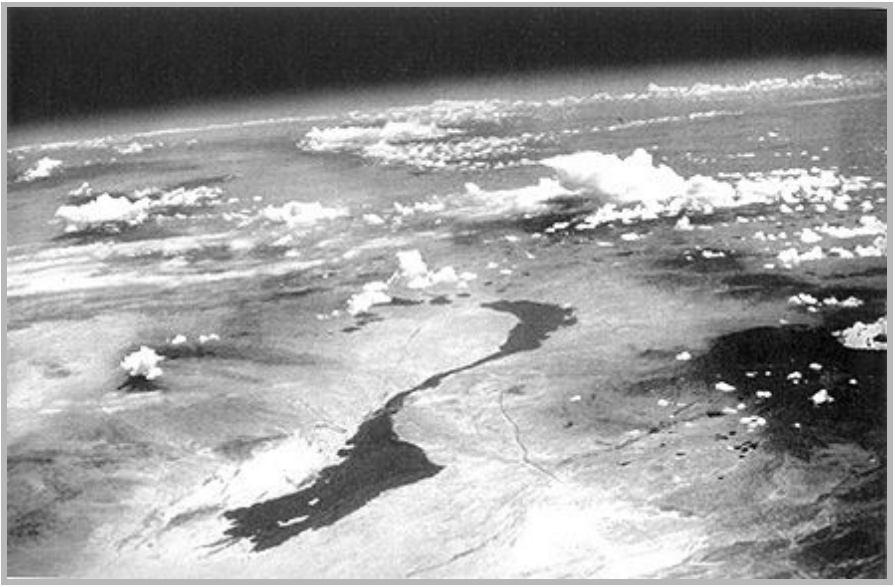

**Figure 1:** One of the first images of the earth from the atmosphere using *V-2* satellites. [4]

advertised NASA as a civilian program, which was intended to promote peaceful and scientific applications of the space. But the underlying motivation behind NASA was clear -protect the national security concerns of the US. Within just two years after *Sputnik-1*, the Soviets achieved a significant lead in space-race by landing the first manmade object to land on the surface of the moon - *Luna-2*. The finish line of the space-race was to land the first humans on moon, a feat that was achieved by the *Apollo-11* mission of NASA.

After the tests on nuclear bomb, no scientific pursuit has had the same level of resources pushed into it than space-race. But in doing so, an unparalleled gain in knowledge was reached in human's millennia of curiosity towards their place in the universe.

## 1.2 From Space-Exploration to Fundamental Science

The goal of both the superpowers during the space-race was to gain hold on space territories that can impact national security. Hence, all the efforts of space-research was diverted to space-exploitation rather than pursuit of fundamental science and curiosity. On the other hand, the field of astronomy until early 1900s was dictated by sensitivity of earth-based telescopes. In 1912, the Austrian-American physicist Victor Hess concluded through balloon experiments a bombardment of heavy particles on the earth's atmosphere that had origins beyond our solar-system (now knows as cosmic-rays). This demonstrated that we can understand the cosmos beyond the use of optical telescopes. Interestingly, during the peak of space-race in 1967, the *Vela* spacecraft of the US, detected an excess of gamma-rays with energies unlike ever seen before from a nuclear test. The first

---

implications of this seemed that the USSR had violated the *Nuclear Test Ban Treaty*, which "prohibits nuclear weapons tests or any other nuclear explosion in the atmosphere, in outer space, and underwater" [5]. A subsequent investigation by the Los Alamos Scientific Laboratory of the data set from the *Vela* satellite revealed the origins of these gamma-rays beyond our galaxy [6]. This was the first detection of gamma-ray bursts, which are the most energetic signals in electromagnetic radiation in the universe. Following this discovery, in 1968, NASA successfully launched *Orbiting Astronomical Observatory*, which became the first satellite in space devoted for astronomy. The development of this satellite carved a new chapter in joining fundamental science and space, with the establishment of a dedicated Astrophysics-wing at NASA that went on to develop the *Great Observatories* program, a result of which is the *Hubble Space Telescope.*

## 1.3 Space Program vs. Space-Science Program

The early intentions of funding space research, as noted in previous sections, was based mainly due to its applications to national security. As the space-based technologies was declassified over time, it formed broad commercial applications to the field of telecommunications. The pursuit of science and exploring the universe, on the other hand, still remains the leading symbolic reasoning for funding space activities.

A space program of a country is usually an amalgam of three sectors: science, commerce and defense. In the context of the US, the National Space Program[7] has specific guidelines for each of these sectors and are divided among various agencies and departments such as NASA, United States Geological Survey (USGS), National Oceanic and Atmospheric Administration (NOAA), United States Trade Representative, Office of Space Commercialization, Secretary of Defense and Director of National Intelligence.

Some of the key highlights that are worth noting from each sector, as quoted in official document of the US, are:

**Space-Science and Exploration (<u>Civil Space Guidelines</u>, NASA):**

- "Set far-reaching exploration milestones. By 2025, begin crewed missions beyond the moon, including sending humans to an asteroid. By the mid-2030s, send humans to orbit Mars and return them safely to Earth."

- "Continue a strong program of space science for observations, research, and analysis of our Sun, solar system, and universe to enhance knowledge of the cosmos, further our understanding of fundamental natural and physical sciences, understand the conditions that may support the development of life, and search for planetary bodies and Earth-like planets in orbit around other stars."

---

[5] Treaty Banning Nuclear Weapon Tests in the Atmosphere, in Outer Space and Under Water, US Depratment of Space, 1963, URL: http://www.state.gov/t/isn/4797.htm

[6] Observations of Gamma-Ray Burst of Cosmic Origins, 1973, URL: http://tinyurl.com/qdamc6x

[7] National Space Policy of The United States, 2010,
URL: https://www.whitehouse.gov/sites/default/files/national_space_policy_6-28-10.pdf

**Environmental Earth Observation & Weather (<u>Civil Space Guidelines</u>, NASA + NOAA):**

- "conduct a program to enhance U.S. global climate change research and sustained monitoring capabilities, advance research into and scientific knowledge of the Earth by accelerating the development of new Earth observing satellites."

**Land Remote Sensing (<u>Civil Space Guidelines</u>, NASA + USGS):**

- "Continue to develop civil applications and information tools based on data collected by Earth observation satellites."

**Commercial Space Guidelines:**

- "Develop governmental space systems only when it is in the national interest and there is no suitable, cost-effective U.S. commercial or, as appropriate, foreign commercial service or system that is or will be available."

**National Security Space Guidelines:**

- "Develop, acquire, and operate space systems and supporting information systems and networks to support U.S. national security and enable defense and intelligence operations during times of peace, crisis, and conflict"

If a similar space research program is drafted by a developing nation, one can argue it should likely focus towards all the above listed areas, except pursuit of space science and exploration. Even NASA over the years has evolved its presence in sectors beyond simply space-science and exploration. In fig. 2, I highlight division of NASA budget for fiscal year 2014. The "science" sector has further division among earth science and astrophysics, with the former focusing on NASA's dedicated research on climate change and land sensing. Thus civilian space program of the US is still largely governed by NASA.

In this study, the author defines a "space-science program" to constitute the three main topics listed above from the <u>Civil Space Guidelines</u> of the US National Space Policy.

## 1.4 Difficulty in Sustaining Space-Science Programs

The primary arguments that are set for discouraging developing nations from having a long-term space-science science program are (a) cost of such programs (b) uncertainty in return of investment.

Even the United States, which has poured more resources towards space-science than any other nation, has ended up substantially dropping its funding for NASA over the years (see fig. 2). From 4.4% percentage of US Budget diverted towards NASA during the peak of space-race in 1966, it has dropped to 0.5% since 2010. Thus, even developed nations with strong economy can find it difficult to sustain committed investment in space-science. The

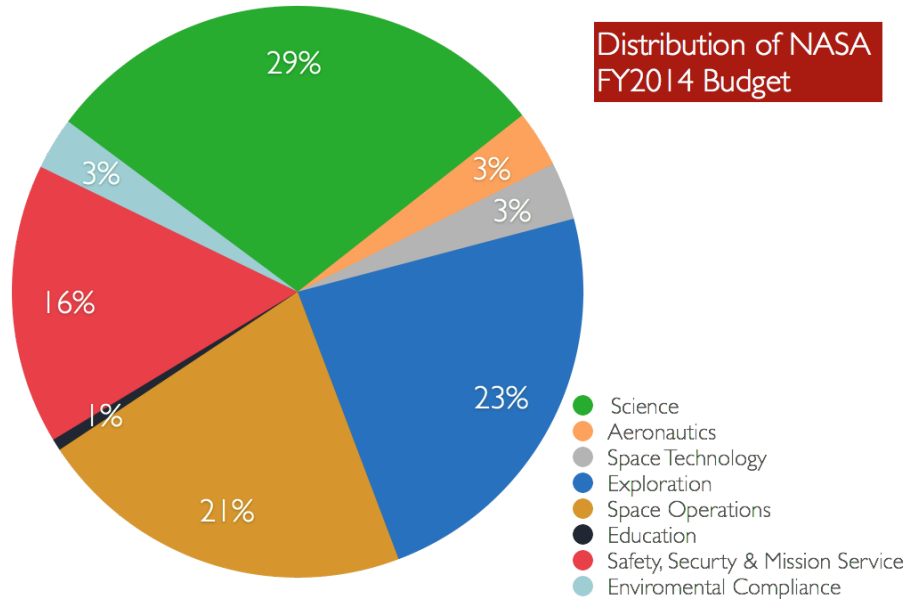

**Figure 2:** Distribution of NASA Budget for FY 2014. The three main areas of focus are science, exploration and mission safety. [8]

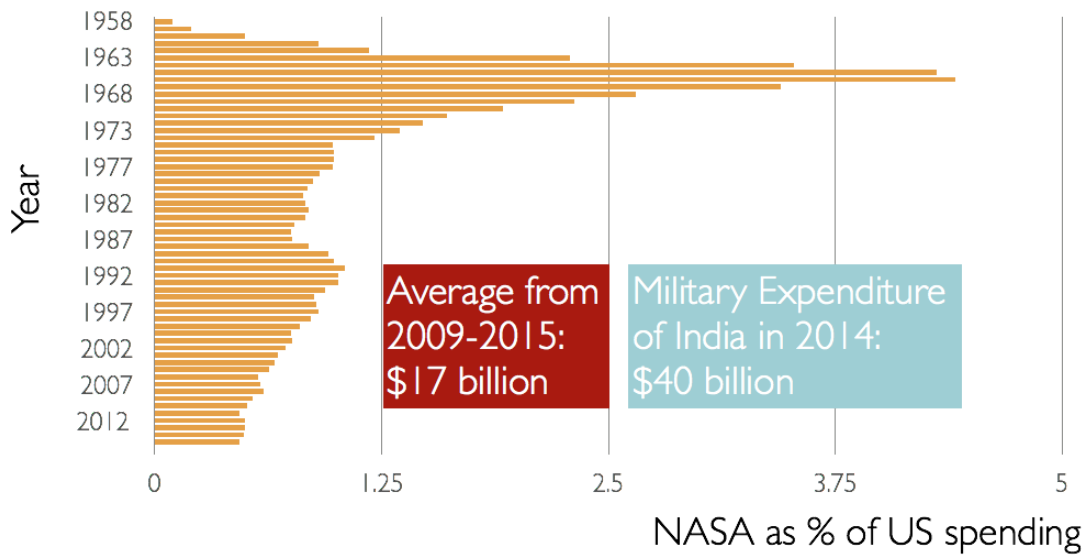

**Figure 3:** Percentage of United States Budget allocated for NASA from 1958-2105. The average annual budget of NASA over last six years is about $20 billion dollars. Compare that with a developing nation such as India, whose year entire military expenditure is only twice that of NASA. However, the ratio of the GDP between the US and India is 0.3. [9]

---

numbers get even more discouraging when we compare the current low-budget phase of NASA for a developing nation, such as India. Due to historic conflicts at its international borders, India is required to spend more money into military expenditure than any other developing nation in the world.[10] Even then, the total military funding of India is only a factor of 2 more than the the current funding of NASA.[11]

For a developing nation to sustain a space-science program, it has to identify set of key areas that directly get benefited because of space research. Unlike NASA, where funding can be spread for ambitious scientific activities (see fig. 2), a developing nation has to pursue space research that has direct overlap with commercial and, for a cases like India, defense benefits. At the same time, to avoid hostility with international community, it cannot simply devote all of its space research towards military applications. Neither can a viable commercial space program sustain without technical knowhow of space-science research.

## 1.5 Emergence of International Cooperation

Any single nation with a matured space-science program would find it very challenging in the current global economy to sustain space-science program in an independent form. Thus, all major nations with space-science research have opened up for international cooperations.

In the context of the United States, the National Space Program keeps international cooperation as one of its principle goals, by stating:

- "Expand international cooperation on mutually beneficial space activities to: broaden and extend the benefits of space; further the peaceful use of space; and enhance collection and partnership in sharing of space-derived information. "

Even China, which has usually maintained secrecy towards its space research, including the budget of its civilian space program, stated in the 2011 "white paper" of its 5 year vision for space that:

- "China maintains that international exchanges and cooperation should be strengthened to promote inclusive space development on the basis of equality and mutual benefit, peaceful utilization and common development."[12]

---

[10] Table 1 of the Stockholm International Peace Research Institute, URL:
http://books.sipri.org/product_info?c_product_id=496#

[11] The author would like to point that the GDP of the US is a factor of 3 times more than that of India
[12] Section 5 of China's Space Activities in 2011, URL:
http://news.xinhuanet.com/english/china/2011-12/29/c_131333479_8.htm

For a space research program from a developing nation, India, was one of the earliest to realize its dependence on international cooperation.  The 2013-14 policy document of ISRO states:

- "Indian Space Research Organisation places great importance on working together with other countries and international bodies in promoting the development and use of space technology for different applications"[13]

# 2.  A Test for International Cooperation in Space

As established in the previous section, an individual nation by itself would generally find it difficult to sustain a space-science program. To encourage a developing nation to invest in space-science, it has to actively engage in diplomacy and international cooperation. But does international cooperation guarantee a significant advancement towards space-science? This section highlights our methodology to test the following hypothesis:

***International cooperation is the cornerstone of a nation's space-science policy***

I test this hypothesis for two classes of case-studies:

- **case-A:** a developing nation which uses international cooperation to advance its space-science program
- **case-B:** a set of nations that collaborate to enhance a common space-science program or objective

In the following subsections I define five variables that measures "success" of a such space-science collaborations. All the measures are ranked in a qualitative manner, based on the supporting literature and relative comparisons between different space programs and their varying socioeconomics. While using our methodology to test case-studies in section 3-4 each of the stated variable below will be ranked in one of the three categories of success-values: (i) low (ii) moderate (iii) significant. **Table-1** is an example of "scorecard" where we record success from each variable for a given case-study.

## 2.1 Scope and Achievement of Science Goals

Whether a country calls their venture a "space program", or "space-science program" or "civilian space research program", independent of all these terminology, the main test of any space activity is whether scientific knowledge is obtained about or using outer-space. As one of the motivations to collaborate with an international partner for space-activity is to implement loftier science goals, this variable will test whether a collaboration was able to push boundaries that otherwise an individual nation may not be able to achieve. In some sense, the choice of this variable is to check if a nation with less matured space-science program is advancing its the technological and scientific knowhow, instead of getting free press and prestige by tagging along with a more stronger space research program. Once the scientific goals are stated the next aspect of this variable is to measure the achievement of those goals.

---

[13]ISRO Annual Report 2013-14, URL:  http://dos.gov.in/rep2014/internationalcooperation.html

In our methodology, "science goals" (short) is counted as an **independent variable** whose success-value solely depends on the space-science program and no other external factors.

## 2.2 Transfer of Technology

Any space-science program has to have technological applications, or in a fancier term, spin offs, to fields beyond space studies. In the context of developing nations, some of such key sectors could be agriculture, meteorological department, telecommunications, disaster management. To measure the effect of international cooperation on such transfer of technology, I will measure if it helped developing nations energize sectors which otherwise would have taken several years of delay (case-A). In the context of when collection of nations are collaborating in space programs, case-B, I will measure if this led to groundbreaking technological feats.

In our methodology, "transfer of technology" is counted as an **independent variable** whose success-value solely depends on the space-science program and no other external factors.

## 2.3 Impact on National Security

As it was the case in the initial development of "civilian" space program such as NASA, it has to build a strategic plan based on the impact on national security. A space-science program in a developing nation may or may not involve directly with defense related research, but one has to see the impact of the technological knowhow towards various aspects of national security. Also, national security from a perspective of space can cover broad areas such as monitoring weather and solar activities, space-debris, applications of expertise in orbital-dynamics towards missile developments. To measure the impact of international cooperation in space-science towards national security, I will test whether nations have involved in formal agreement towards of space-security and whether they have engaged in civilian space-science missions that can be shifted, during an emergency or crisis, towards protecting national interests.

In our methodology, "national security" is counted as an **dependent variable** whose success-value can depend on many external factors beyond the space-science program.

## 2.4 Impact on Economy

To develop and sustain a successful space-science program, even in collaboration with other nations, a developing nation has to pour significant amount of its GDP towards space research. But at the same time, if nation develops its own launch capabilities and active geosynchronous and remote sensing satellites, they can use this for many commercial applications. To measure the impact of international cooperation on the commercial aspects of space-science programs, I will test whether nations have signed official agreement with a space program.

In our methodology, "economy" is counted as an **dependent variable** whose success-value can depend on many external factors beyond the space-science program.

## 2.5 Impact on Science Education

Promoting science, technology, engineering and mathematics (STEM) education using space-science programs could prove an expensive affair for developing nation. There could be many ways a nation can go about increase STEM graduates than building space-museums that highlight research conducted solely in their nation! But one measurement of a space-science towards education could be the overall impact on research in the fields of science and technology. If one looks it context of a developing nation, the impact of international cooperation on this particular variable is of second-order, i.e. it is impacted more due to the fact here is a space-science program which could be enhancing due to international cooperation. However, if one looks it in the context of a collaboration among nations, this can have a more of one-to-one map by measuring research impact and number of publications that have resulted mainly because of such collaboration.

Because of varying impact of this parameter between cases A and case B, and that the main goal of this paper is to context of developing nations, "science education" is counted as an **dependent variable** whose success-value can depend on many external factors beyond just the space-science program.

## 2.6 Understanding of The Universe

From an idealist standpoint, the ultimate goal of a space-science mission is to advance mankind's understanding of the about the formation and evolution of the universe. This criteria of course can get very subjective - does a satellite mission dedicated for remote sensing enhance our knowledge of the universe? It does teach us about geology that may have impact in our understanding of planets beyond solar-system, but that is not a direct goal and relies on many uncertainties. So a more focused definition of this variable is impact towards the field of astrophysics. And to make it even narrower, space exploration by itself does not directly impact astrophysics. By astrophysics I mean a direct impact towards fundamental science from space-based missions. One clear example of this definition is the *Hubble Space Telescope*. In the context of both case-A and case-B, I measure if such collaborations have engaged and enhanced the astrophysical goals.

In our methodology, "astrophysics" (short) is counted as an **independent variable** whose success-value solely depends on the space-science program and no other external factors.

## 2.7  Framework of Quantifiable Variables

To test particular examples using these six variables, in the upcoming Section 3-4 and I use the following case-studies:

(i) Space research program of India (example of case-A)

(ii)  International Space Station (example of case-B)

| Criteria / Success | Low | Moderate | Significant |
|---|---|---|---|
| Science Goals | | | |
| Transfer of Technology | | | |
| National Security | | | |
| Economy | | | |
| Science Education | | | |
| Astrophysics | | | |

**Table 1:** A sample scorecard to measure success-value of each variables for a given case study. For each test case, I will record impact of each of these variables by quoting 'x' in one of the columns.

# 3. <u>Case-A:</u> Indian Space Research Organization

*"Very many individuals with myopic vision questioned the relevance of space activities in a newly independent nation, which was finding it difficult to feed its population    The vision of the founders was clear: if Indians were to play meaningful role in the community of nations, they must be second to none in the application of advanced technologies to their real-life problems. **They had no intention of using it as a means to display our might.**"*

<div align="right">- Dr. A. P . J Abdul Kalam, 11th President of India</div>

## 3.1 Overview

A formal space research program was established in India as early 1962, arguably by the influence of the space-race between the superpowers. Within a month of *Apollo 11's* moon landing in 1969, India initiated a dedicated civilian space program by establishing the Indian Space Research Organization (ISRO). To a reader not familiar with the developments of ISRO, fig. 3 showcases a timeline of major developments.

The Indian space program is unique in many ways. If we compare the the top nine[14] most funded civilian space programs in the world (see fig. 3), the first thing one can notice is the difference between United States and rest of the world when it comes to space-science. The total budget NASA is more than the next 8 space programs combined, making it an undisputed leader in the field of space-science for years to come. The only space programs that comes close to NASA are the *Russian Federal Space Agency* and the *European Space Agency* (both agencies have funding almost 3 times less than that of NASA).

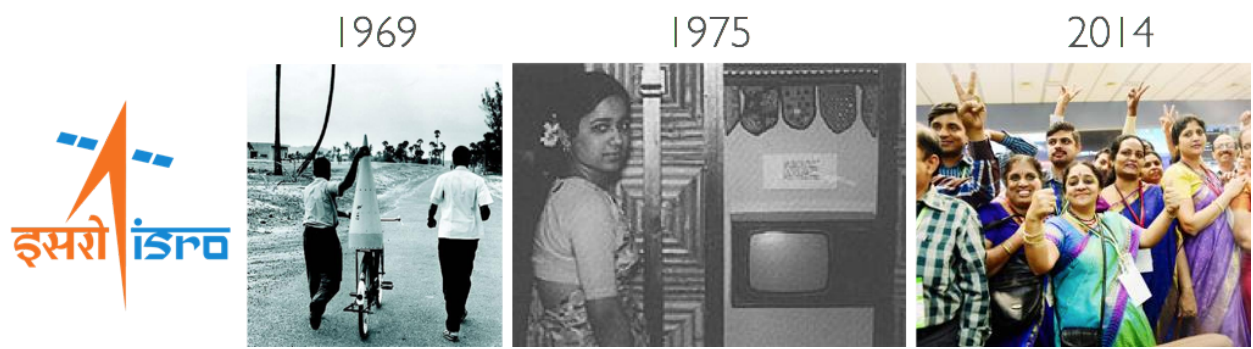

**Figure 3:** from left-right: (i) logo of ISRO (ii) early stages of missile testing (iii) a ISRO technician operating India's first telecommunication satellite (iii) ISRO scientists celebrate after the success of Mars mission

---

[14] The 10th most space funding seems to be from Ukraine but the author could not find any credible source for the exact amount spent on civilian space program.

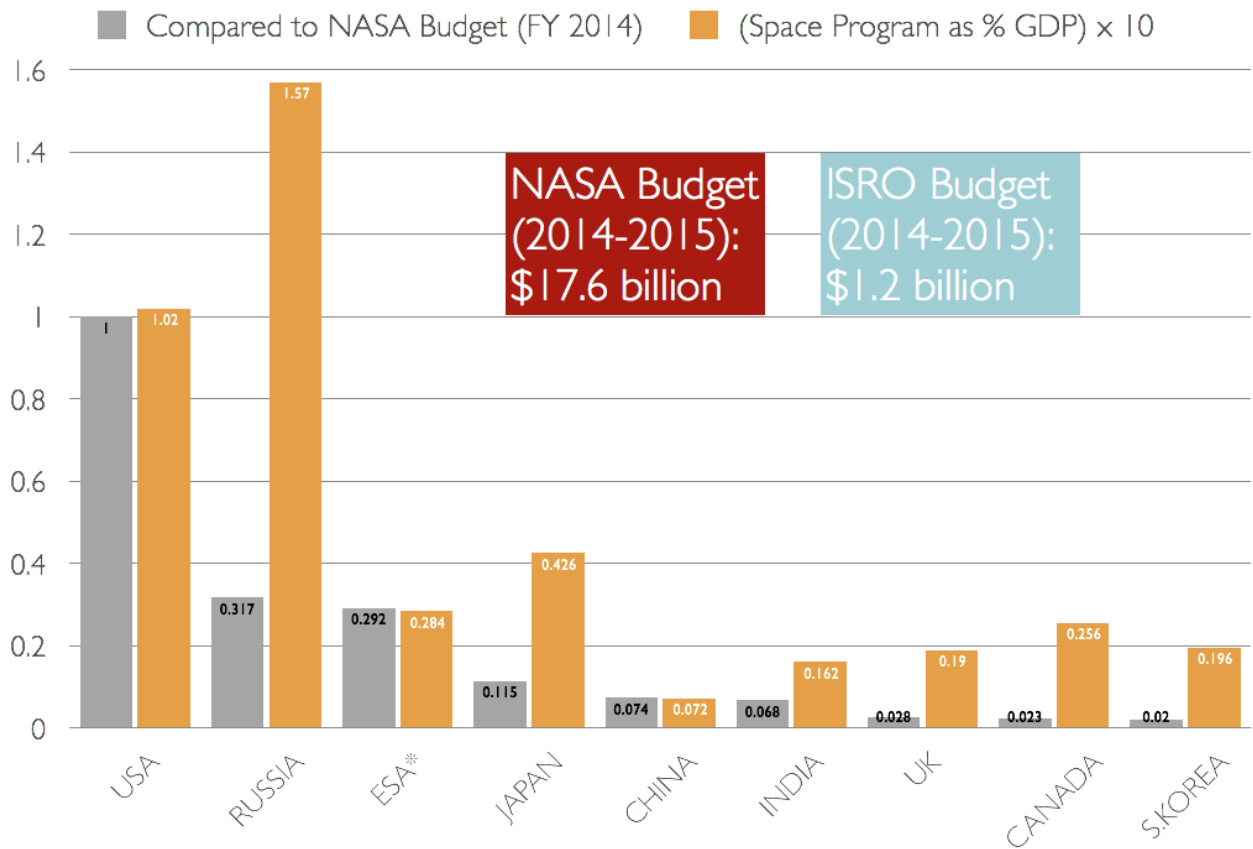

**Figure 4:** Comparison of the top nine civilian space programs in the world. The grey bars makes relative comparison between NASA and funding from other space agencies. Hence the measure of this for the USA would be 1. The yellow bars measures the percentage of GDP poured into civilian space research in each counties. For the European Space Agency, which has 22 member european nations, I have used the GDP measurements of europe. [15] [16] [17] [18] [19] [20] [21] [22] [23] [24]

The other interesting thing to note about fig 4. is that the only developing nation to be featured in this list of highly funded space programs is India. If one compares purely in terms of total funding, the budget of ISRO is 15 times less than the total budget allocated for NASA. It is still the sixth most funded civilian space program, lagging behind China's total civilian space funding by only $9 million.  Also, barring China, India and South Korea, all of the stated countries in fig. 4 are the initial funders for International Space Station.

In a global economy where even the funding of NASA has dropped in recent times (see fig. 2),  India not only is able to surpass majority of economically developed nation in terms of its space budget, but it has consistently kept increasing the civilian space funding since 2009 (see fig. 5), after the successful mission to moon. In doing so it achieved the scientific feat of being the first Asian nation to reach Mars in 2013. If NASA is labeled as the ideal example of space-science programs that other nations should collaborate with, arguably it is ISRO that stands as a role model for the developing nations to rethink their policy about space research.

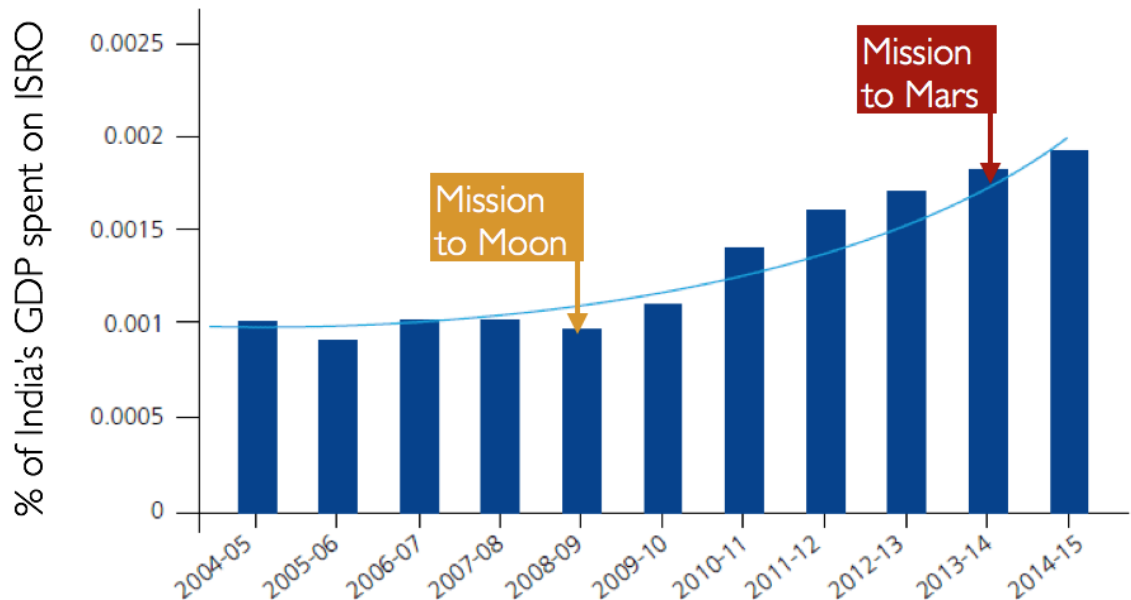

**Figure 5:** Percentage India's GDP devoted to space science research from 2004-2015. [25]

---

## 3.1 Engagements with International Community

*"Over the years, as ISRO has matured in experience and technological capabilities, the scope for cooperation has become multifaceted. While exploratory missions beyond the earth are the natural candidates for such cooperative efforts, there are many other themes like climate change on earth that are of interest to international cooperation because of their global impact."*
- ISRO, Policy Document on International Cooperation [26]

If one tracks the major development of ISRO over the years, there is almost always an international partner that has provided the key support. From the launch of its first satellite *Aryabhatta* from the USSR, which interestingly was dedicated for astrophysics, to collaborations with NASA on dedicated educational satellites starting as early as 1975, ISRO has strengthen international cooperation with peaceful and science oriented space activities. As of 2014, ISRO has formal agreements with 33 nations in utilizing applications of civilian space research.

A particularly important international partnership Indian space program shares is with the US. The 2014 vision statement of the US-India Strategic Partnership highlighted this aspect:

"Joint research and collaboration in every aspect—ranging from particles of creation to outer space -- will produce boundless innovation and high technology collaboration that changes our lives". [27]

One of the immediate effect of this strategic partnership was a new level of collaboration between NASA and ISRO. Both agencies have signed on a joint Mars Working Group that will enhance long term exploration of Mars. [28]

---

[26] ISRO Policy Document on Intenrational Cooperation, URL: http://www.isro.gov.in/international-cooperation

[27] Joint US Indian Strategic Partenrship, 2014, URL: https://www.whitehouse.gov/the-press-office/2014/09/29/vision-statement-us-india-strategic-partnership-chalein-saath-saath-forw

[28] US-India to collaborate on Mars, NASA, 2014, URL: https://www.nasa.gov/press/2014/september/us-india-to-collaborate-on-mars-exploration-earth-observing-mission/#.Vk3PDN-rSRs

## 3.2 Testing the Impact of International Cooperations

As noted so far, ISRO has stood out against all claims that discourage developing nations from sustaining ambitious space-science program. In doing so ISRO provides a unique case study to test whether its success has resulted from its willingness for international cooperation. Using the methodology defined in Section 2, I rank the success-value for the six variables in the case of ISRO.

### 3.2.1 Science Goals

The two most important science missions conducted by ISRO are,
(i) *Chandrayaan*-1- lunar probe in 2008
(ii) *Mangalyaan-1* - mission to Mars in 2014

Both by itself set quite a milestone for a space-program from developing nations to attain. What is particularly important to note is the kind of unique science these missions were able to extract using science payloads from ISRO's international partner. In Chandrayaan-1, there were 6 science payloads from NASA, ESA and the Bulgarian Space Academy that was given for free of cost to ISRO. One of the NASA payload help to investigate ice formation in the polar caps of moon. During about one year of lifetime, India's maiden lunar probe achieved upto 95% of its science goals

The launch of Mars Orbiter Mission, Mangalyaan-1, was carried from India in November of 2013. Prior to the launch, ISRO had an agreement with NASA to utilize the Deep Space Network Facilities to monitor the probe after the launch.[29] However, during the same time there was a government shutdown in the US. In a rare symbolic gesture to promote science, NASA and JPL reaffirmed their support to ISRO and monitored the launch even during shutdown.

Using the cited examples, I assess the success-value of international cooperation for the variable "science-goals" as significant.

### 3.2.2 Transfer of Technology

Over the years, Indian space program has generated a wide range of technology transfer[30] and spin-offs[31] to enhance to climate predictions for important sectors like meteorology and agriculture.  In the context of the international cooperation, the early impetus resulted with collaborations with the US, France and Germany during 1970s.

---

[29] IRSO Press Release, October 2013, URL:
http://web.archive.org/web/20131017141915/http://www.isro.org/pressrelease/scripts/pressreleasein.aspx?Oct05_2013
[30] ISRO Technology Transfer Portal, URL:
http://www.isro.gov.in/isro-technology-transfer/technologies-transferred

[31] ISRO Space Spin-off Portal, URL:
http://www.isro.gov.in/isro-technology-transfer/space-spin-offs-isro

In 1975, ISRO in collaboration with NASA conducted Satellite Instructional Television Experiment (SITE)[32]. Using this unique satellite experiment, the Government of India was able telecast educational programs to over 2400 villages. Farmers were provided specific guidance to adapt over seasonal changes through this program.  To think about this feat in today's context of internet connectivity might trivialize the judgement, but during this time India barely had telephone connectivity beyond urban parts.

With success of SITE, Indian space research made its first attempt on the telecommunication satellite and in 1977 initiated the Satellite Telecommunications Experiments Projects (STEP) from Franco-German satellite *Symphonie*. These initial experiments gave ISRO the technical knowhow to develop its own satellite systems. From 1980s onwards, ISRO developed  *Indian National Satellite System (INSAT)* which were dedicated towards meteorology, telecommunications and broadcasting,

Using this historical developments, I assess the success-value of international cooperation for the variable "transfer of technology"  as significant.

### 3.2.3  National Security

Until very recently, ISRO did not actively collaborate with India's Defense Research and Development Organization (DRDO) . In 2013, ISRO launched *INSAT-4F* which is the first dedicated satellite from India to have application to defense. This satellite help with marine communications for the Indian Navy. It is interesting to note that even though India has its own launch capabilities, this satellite was launched from the space-agency of France.

One of the important outcomes of the renewed US-India Partnership was the formation of US-India Space Security Cooperation. The joint statement emphasized:

"U.S.-India civil cooperation in space has not led to extensive cooperation on space security, at least to date… this is a time of transformation and progress for our strategic partnership, so too is it a time of growth for our space security relationship. Our governments recognize the importance of space security." [33]

Barring these recent developments, I conclude a general delay from international partners towards enhancing India's space defense capabilities. Thus, the success-value of international cooperation for the variable "national security"  is measured as moderate.

---

[32] Space Application Center,  ISRO, URL: http://www.sac.gov.in/SACSITE/asac-history.html

[33] U.S.-India Space Security Cooperation: A Partnership for the 21st Century, Department of Space, 2015, URL: http://www.state.gov/t/avc/rls/2015/238609.htm

### 3.2.4  Economy

One important economic criteria about ISRO compared to other major space programs is the cost-effectiveness of its mission. The commercial arm of India's space program - Antrix- leads the effort of signing agreements with nations and private corporations to use satellite and launch facilities from ISRO. As of 2014, India has launched 45 satellites for 19 different nations. Between 2015-2017, India will be launching another 28 satellites of Algeria, Canada, Germany, Indonesia, Singapore and the US.[34]

This has provided by India a significant return of investment from its space activities. One estimate from the India's Department of Space quoted that the income generated from such commercial satellite launch has added "to around USD 17 million and Euro 78.5 million" to the economy.

Citing this figures, I assess the success-value of international cooperation towards India's space program from "economic" perspective as significant.

### 3.2.6  Science Education

One common perception is that a nation's funding of space-activities inspires generations of students to research in STEM. In context of India, if we measure the impact of science education based on impact of research in STEM, the signs are rather disappointing and points towards a nil effect. As shown in fig. 6, the "scholarly citation impact" of India from 2001-2013 has remained fairly flat and much below the world-average. Notice for this same time period there has been a steady increase in the budget of ISRO (see fig. 5).

Attempts have been made to accelerate the impact on international cooperations in space to enhance science education. The Joint Civil Space Working Group between US and India met in 2011 to discuss India's participation in NASA-led Global Learning and Observations to Benefit the Environment (GLOBE) education program.[35] But the total schools from India enrolled in this program is about 1000, which can make very little progress towards India's complex science education woes.

Citing this figures and examples, I assess the success-value of international cooperation and ISRO towards "science education" as low.

---

[34] Report from The Economic Times, 2014:
http://economictimes.indiatimes.com/news/science/india-to-bear-cost-of-building-launching-of-saarc-satellite-says-jitendra-singh/articleshow/48174416.cms

[35] US-India Space Cooperation, 2011, URL:
https://www.whitehouse.gov/sites/default/files/india-factsheets/Fact_Sheet_on_U.S.-India_Space_Cooperation.pdf

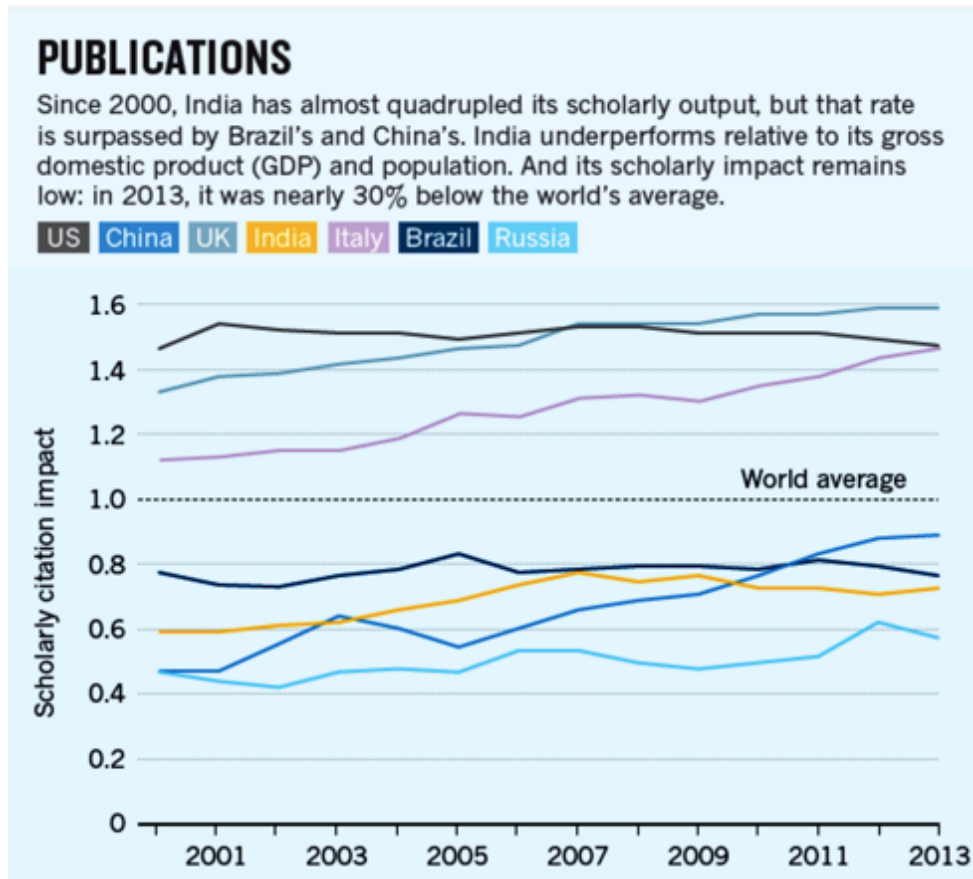

**Figure 6:** Impact factor of publications from US, China, UK, India, Italy, Brazil and Russia during 2000 to 2013. [36]

### 3.2.6 Astrophysics

The first space satellite launched by ISRO in 1975, *Aryabhata* (named in honor of ancient Indian astronomer-mathematician who invented zero and trigonometry), was dedicated towards x-ray astronomy and solar physics.[37] But barring a few forgettable cases, ISRO has not actively pushed satellites for astrophysics. In 1996 it launched a satellite for X-ray astronomy but had limited scientific impact.

The most ambitious astrophysics satellite from India towards astrophysics and cosmology, ASTROSAT - a dedicated multi-wavelength satellite, was launched as recent as September

---

2015.[38]  When one compares this with the timeline of astrophysics mission from NASA and ESA, it seems the Indian space agency did not earn a great deal towards enhancing fundamental science. This could also be a reason for low science education value as stated in for previous variable. Thus, I conclude the success-value of international cooperation towards astrophysics to be low.

## 3.3 Rankings from Framework for ISRO

| Criteria / Success | Low | Moderate | Significant |
|---|---|---|---|
| Science Goals | | | X |
| Transfer of Technology | | | X |
| National Security | | X | |
| Economy | | | X |
| Science Education | X | | |
| Astrophysics | X | | |

**Table 2:** Scorecard of ISRO.

# 4. <u>Case-B:</u> International Space Station

*"Our progress in space, taking giant steps for all mankind, is a tribute to American teamwork and excellence..  And we can be proud to say: We are first; we are the best; and we are so because we're free…Tonight, I am directing NASA to develop a permanently manned space station and to do it within a decade.. NASA will invite other countries to participate so we can strengthen peace, build prosperity, and expand freedom for all who share our goals."* [39]

- State of the Union Address, President Ronald Reagan, 1984

## 4.1 Overview

International Space Station is the most ambitious attempt of human race towards space exploration. It will also go down in history as the costliest international cooperation since the formation of United Nations. Since its launch in 1998, the total money poured into International Space Station (ISS), based on some ballpark estimates comes to about $200 billion. [40] The next most expensive scientific instrument after ISS is the Large Hadron Collider, which costed about $10 billion. In fact, one fifth on NASA's current total budget is still devoted towards to maintain ISS.

In terms of scale, the ability of international community to put this level of infrastructure in space is unprecedented. To construct this 419455 kg of structure, it took 115 space missions over the years. ISS flies in the low-earth orbit, which is about 400 km from the surface, and takes approximately 90 minutes to finish one orbit around the earth.[41]

## 4.2 Historic role of diplomacy

When President Reagan announced the plan of building ISS, he made explicitly clear the role US allies were to play in the development stage. The original partners of ISS were Japan, Canada and 11 participating nations of the European Space Agency. Although the cold war was well over and lessons were learnt from the space-race, the US did not initially want USSR to be part of the ISS. But with the change in diplomatic relations since the collapse of Soviet Union and rising estimated cost of ISS construction, it became clear Russia could play a significant role.  The original budget assigned to NASA for this mission was $8 billion. By the time of launch the total number had reached $60 billion.

Also, unlike NASA, Russia also had significant expertise in building and maintaining space-stations. During the Clinton administration, Russia formally signed to be the partner

---

in the ISS. Interestingly, the proposed name for ISS until then was *Freedom*, likely inspired by Reagan's State of the Union address. But that name was dropped and both Russia and US agrees simply on a "space-station". The official agreement between the 15 participating nations was signed on January 28, 1988. At the mark of the 15th anniversary of ISS, both Russia and US signed to keep the station functioning till at least 2024.[42]

## 4.2 Testing the Impact of International Cooperations

As ISS is the first of a kind of international cooperation for space infrastructure, it provides unique case study to test the methodology described in Section 2.

### 4.2.1 Science Goal

The main science goal of the ISS was to successfully and safely construct it in first place. Although there are set of dedicated micro-gravity experiments that were carried successfully in variety of fields like biomedical research, biotechnology, physical sciences and material science, most of the science goals evolved over time based on collaboration. In fig 7, I highlight the total scientific collaboration that were led betweens scientists from the five participating space-agencies.

As each aspect of science goal crucially depended on international cooperation between the 15 participating nations, I assess the success-value for the variable "science-goals" as significant.

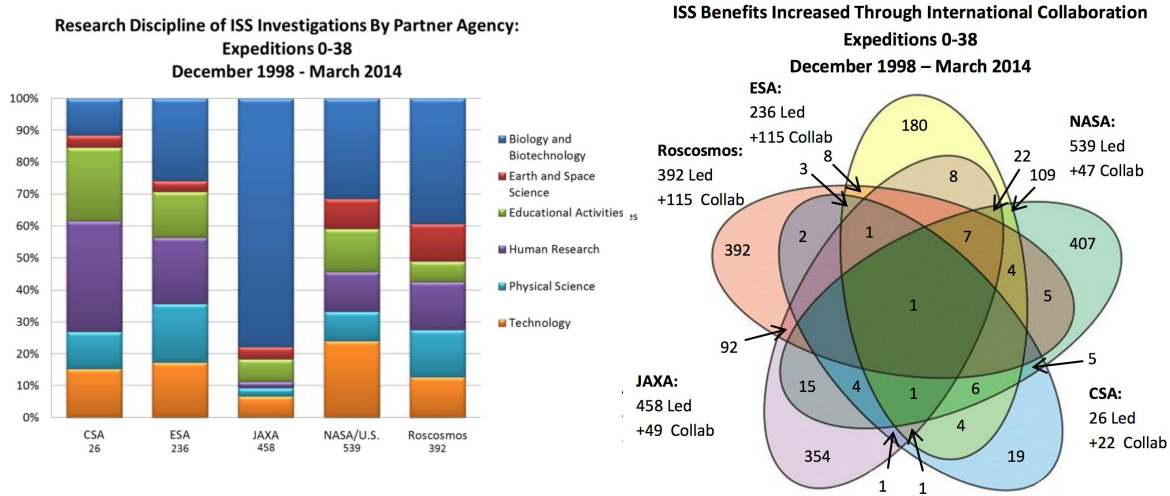

**Figure 7:** Science collaboration between participating nations of ISS. On the left is comparison of each space agency with field of research. On right is collaboration between space programs. [43]

---

### 4.2.2  Transfer of Technology

From the literature, it is not clear whether transfer of technology from ISS was well listed before the mission took place. But over the years the biggest support towards ISS has resulted due various applications of its technologies. Some standout examples are robotics that are used for surgical capabilities and image capture devices. [44]  In the sector of climate prediction, the ISS carries dedicated science payloads that monitor earth's activities. [45]

Based on this examples, we assess the success-value of ISS for the variable "transfer of technology" as significant.

### 4.2.3  National Security

Interestingly for a space mission which took more resources than most defense research projects, there was no explicitly stated application of ISS to national security. One can argue that the inclusion of Russia in ISS prohibited the rest of the NATO allies to put any probe that could conduct deep earth surveillance.  The dependence of the US on Russia, in fact, increased since the *Columbia disaster*. All the current and planned NASA astronauts have to now depend on Russian launch facilities. This put US on disadvantage towards diplomatic pressure on Russia. [46]

Based on this, I can conclude that none of the participating nations of ISS gained any major applications to national security. Thus, the success-value for the variable "national security" is chosen to be low.

### 4.2.4 Economy

It is very difficult to find any positive outcome to economy of any of the participating nations due to their involvement in ISS. The current budget of both NASA and ESA still has significant resources diverted towards ISS.  With almost no monetary return of investment of all 15 nations since its participation in ISS, I rank the success-value for the variable "economy" as low.

---

[44] NASA, 2009, URL:  https://spinoff.nasa.gov/pdf/ISS_Flyer.pdf

[45] NASA, 2012, URL:  http://www.nasa.gov/mission_pages/station/research/benefits/Climate_Change.html

[46] Universe Today, 2014, URL:
http://www.universetoday.com/110006/iss-nasa-and-us-national-security-dependent-on-russian-ukrainian-rocketry-amidst-crimean-crisis/

### 4.2.5  Science Education

A lot of interesting research in STEM took place in the member nations because of the availability of data from variety of science experiments on ISS. Using a modified criteria of judging science education based on research publication, instead of research impact as was the case while comparing ISRO, I list the breakdown based on the research fields in fig 8. The field of biology and human research has probably had the most benefits from ISS. However, even after almost 16 years of resources being poured into ISS, the amount of publications in journals and conferences are only about 1400. This is quite a low number compared to the total number of publications in the US for the year of 2013, which is about 450,000.[47]

Citing these figures, we conclude the impact of ISS on the variable "science education" is moderate.

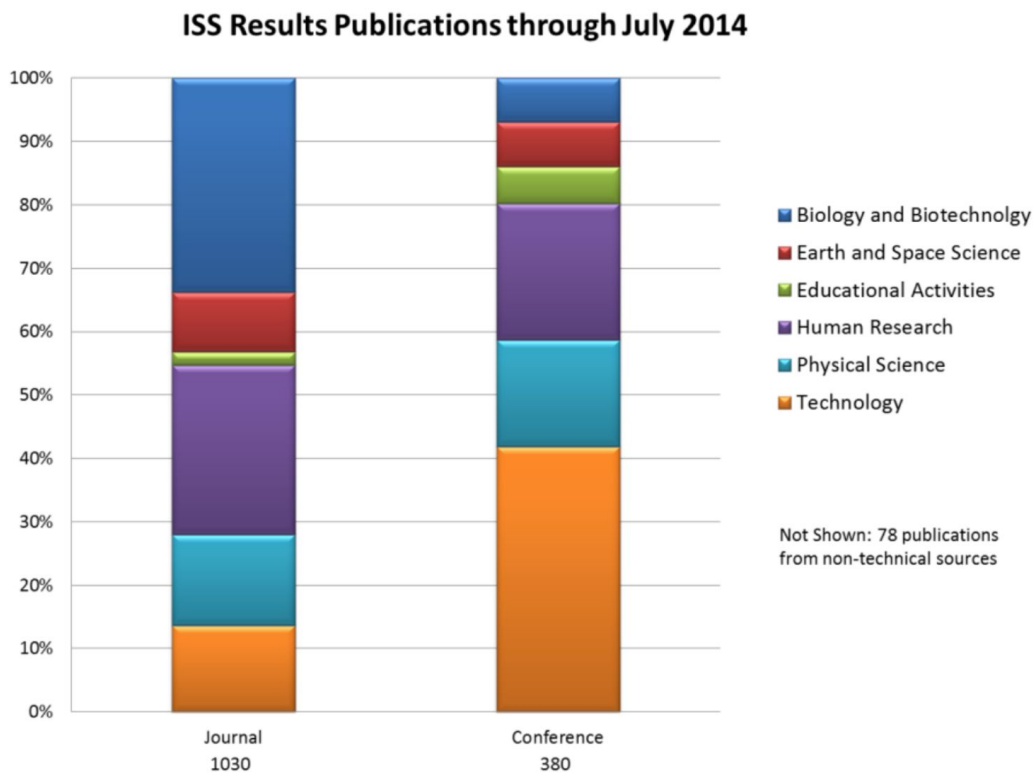

**Figure 8:** Scientific publications resulted from ISS during 1998-2014[48]

---

## 4.2.6 Astrophysics

*"The International Space Station is an orbital turkey.. No important science has come out of it. I could almost say no science has come out of it."*

<div align="right">- Steven Weinberg, recipient of 1979 Nobel Prize in Physics[49]</div>

If one compares the implication of astrophysics, or more poetically towards the understanding of the universe, then ISS is far behind much cheaper space missions like *Hubble Space Telescope*. ISS had the potential to contribute to fundamental physics, like measuring curvature of spacetime around earth from Einstein's general relativity. But the space-station floats in low earth orbit, so it still has substantial gradient of gravitational field to carry sensitive microgravity experiments. The only significant astrophysics module on ISS is the Alpha Magnetic Spectrometer, which was added in 2011. AMS is a cosmic-ray detector and the goal is to better understand formation of dark-matter. Though no new insights have been gained by this $2 billion installment that is not known from earth ground based detectors.[50]

Based on this comparison, I can conclude that none of the participating nations of ISS benefited towards any new understanding of the universe. Thus, the success-value for the variable "astrophysics" is chosen to be low.

## 4.3 Rankings from Framework for ISS

| Criteria / Success | Low | Moderate | Significant |
|---|---|---|---|
| Science Goals | | | X |
| Transfer of Technology | | | X |
| National Security | X | | |
| Economy | X | | |
| Science Education | | X | |
| Astrophysics | X | | |

**Table 3:** Scorecard for ISS

---

[49] Space.Com, 2007, URL: http://www.space.com/4357-nobel-laureate-disses-nasa-manned-spaceflight.html

[50] NASA, 201, URL: http://ams.nasa.gov/

# 5. Limitations of International Cooperations

## 5.1 Chinese Exclusion Policy of NASA

"None of the funds made available for the National Aeronautics and Space Administration (NASA)..to develop, design, plan, promulgate, implement, or execute a bilateral policy, program, order, or contract of any kind to participate, collaborate, or coordinate bilaterally in any way with China or any Chinese-owned company"[51]

- Section 539 of the Consolidated and Further Continuing Appropriations Act of the US, 2012

The China National Space Administration (CNSA) is the fifth most funded civilian space program in the world (see fig. 4). If official collaboration were taking place between NASA and CNSA, a lot of success could have achieved towards long-term space infrastructure. But diplomatic tensions, citing national security concerns over joint space venture, has prevented the two leading space agencies to collaborate. China has been prevented from even multilateral space ventures like the ISS. In case of such international hostility, China has planned its independent space-station to be fully operational by 2022. [52] It is almost unclear how a single nation can sustain cost of such mission. But if China adheres to its principle of international cooperation, one would see Asian nations rallying behind space prowess of China, causing a further dip in the national security concerns of the US.

## 5.2 SAARC Satellite

In a major boost to space diplomacy, India announced a first dedicated satellite for South Asian Association for Regional Cooperation (SAARC) region[53]. The satellite would be dedicated enhance communication and meteorology sectors. India, in a show of rare hospitality, decided to bare all the cost of the satellite.

Although all the member nations supported India's proposal, but Pakistan raised serious concerns over the national security implications of this satellite. Pakistan also questioned regarding the ownership of this satellite and the usage of the data. India has decided to go ahead with the satellite without approval of Pakistan and this is going to raise severe diplomatic issues after the launch of the satellite.

---

[51] Act, URL  http://www.gpo.gov/fdsys/pkg/PLAW-112publ55/html/PLAW-112publ55.htm

[52] Space.com, 2014, URL: http://www.space.com/27440-china-space-station-plans.html

[53] Economic Times, 2014, URL: http://economictimes.indiatimes.com/news/science/pakistan-raises-security-issues-to-oppose-saarc-satellite-project/articleshow/47822734.cms

# 4. Conclusions & Future Work

In today's age, a space science program of a nation, like many of its other sectors, has a global presence. Instead of judging space-science program of nation through conventional measures of prestige or extending mankind's knowledge about the universe, we provide a framework to measure impact on board range of criterias. The six variables stated in my criteria framework can be used as yardstick to measure success of space programs of developing nations and international collaboration.

Using my framework for a space program of developing nation like India, I conclude it has given a boost in achieving space-science goals, economy and transfer of technology to India's important sectors. My framework suggest that a developing nation like India gained very little in terms of applications space research to national security and science education.

Applying the same framework in context of ambitious international collaboration for International Space Station, I conclude it helps achieve scientific goals that otherwise no nation could have scaled on its own. But the value of investment to sectors like economy and national security have proved to be fairly low from such collaborations.

For future studies, I propose to extend this existing framework by having a more stricter division of "science-goals" into criteria such as launch capabilities, space-exploration and satellite development.

The results of existing framework should also be studied in context of the space science program of South Korea, as the country recently promoted to being a "developed nation" and to larger international collaboration such as the European Space Agency.

# 5. Acknowledgements

The author will like to thank Margaret Kosal, Seymour Goodman, Deirdre Shoemaker and the *Sam Nunn Security Program* fellows for constructive feedback on this study. The author will like to thank Narayanan M Komerath for valuable discussions that shaped this study.